\documentclass[12pt]{article}

\newcommand{\sect}[1]{\setcounter{equation}{0}\section{#1}}

\usepackage{latexsym}
\usepackage{amsfonts}
\usepackage{cite}
\usepackage[active]{srcltx}

\def\bra#1{\langle~#1~|}
\def\ket#1{|~#1~\rangle}

\def\Seff{S_{\mathrm{eff}}}

\def\spazio#1{\vrule height#1em width0em depth#1em}
\def\gb{\gamma_{{}_\mathrm{B}}}
\def\ge{\gamma_{{}_\mathrm{E}}}

\date {}
\begin{document}
\bigskip\bigskip
\title{\bf Effective action for fermions with anomalous magnetic
moment from Foldy-Wouthuysen transformation.}
\bigskip\bigskip
\author{A. Barducci and R. Giachetti}

\maketitle \centerline{{  Department of Physics, University of
Florence and I.N.F.N. Sezione di Firenze }}\centerline{{ Via G.
Sansone 1, I-50019 Sesto Fiorentino, Firenze , Italy\footnote{
e-address: barducci@fi.infn.it, giachetti@fi.infn.it }}}
\bigskip
\bigskip

%\bigskip
\noindent

\begin{abstract}
%{{\ccc Abstract.}
In this paper we calculate the effective action for neutral particles with anomalous magnetic moment in an external
magnetic and electric field. We show that we can take advantage from the Foldy Wouthuysen transformation for such
systems, determined in our previous works: indeed, by this transformation we have explicitly evaluated the diagonalized
Hamiltonian, allowing to present a closed form for the corresponding effective action and for the partition function at
finite temperature from which the thermodynamical potentials can be calculated.

\medskip
\noindent
PACS numbers: 03.65.Pm, 11.10.Wx 
\end{abstract}

\bigskip
\bigskip

\sect{Introduction.}

In some recent works \cite{BGJP,BGPJP,BGPEP} we have used Grassmann variables in order to investigate the properties
of the interactions of pseudo-classical spinning relativistic particles and superparticles \cite{BCL1,BM,BDZDVH,BDVH}
with electromagnetic couplings. The description of spinning particles was generalized by allowing for the presence of an
anomalous magnetic moment (in the following: a.m.m.) \cite{B,GS1,GS2,BBC2001}
and the Foldy-Wouthuysen transformation (hereafter FWT) was fruitfully applied to a pseudo-classical spinning particle
with a.m.m. in a classical stationary electromagnetic field \cite{BGPJP,EK,pch04,pch05,pcj05}.
We studied, in particular, the cases in which the result could be expressed in a closed form. These
turned out to be the systems of  a neutral particle in a stationary electric field, a neutral particle in a 
stationary magnetic field and a charged particle in a stationary magnetic field.
An important technical point that made it possible to obtain the results was the possibility of
exploiting two different representations of the Clifford algebra realizing the Dirac brackets 
needed for the quantization of the pseudo-classical variables and intertwined by a Pauli-Gursey unitary transformation, 
that allow us to get more easily the exact form for the three different interacting cases.
Finally, in a later work \cite{BGPEP}, a Schwinger proper time description of these systems has also been given.

The purpose of this work is to study the finite temperature effective action of neutral spinning relativistic particles
with a.m.m. in external stationary uniform magnetic and electric fields. Using our previous
results on the FWT, we will determine the effective action and the corresponding thermodynamical potentials in closed
form. 
\bigskip

\sect{The method.}

In order to illustrate the method we recall that the partition function for a spinor particle in an electromagnetic 
field can be written in the functional integral form as
\begin{eqnarray}
 Z[A,\eta,\bar{\eta}]\!\!\!\!&=&\!\!\!\!\!\int \mathcal{D}[\bar\psi(x)]\,\mathcal{D}[\psi(x)]\,\,
\exp\Bigl\{\,i\int\!\! d^4x\Bigl[ \bar\psi(x) \bigl(i\hat\partial     -q\hat A(x) -m\bigr)
+\bar\eta(x)\psi(x)+\bar\psi(x)\eta(x)   \Bigr]   \Bigr\}\spazio{1.4}\cr
 \!\!\!\!&=&\!\!\!\!\!\,\,\exp\Bigl\{\,-i\int\!\! d^4x\int\!\! d^4y \,\,\, \bar\eta(x) S_F(x,y)\eta(y)   
\Bigr\}\,\!\cdot\!\, \det\bigl[ -i S_F^{-1}\bigr]
\label{Z}
\end{eqnarray}
where $\bar\eta(x),\,\eta(x)$ are the fermionic sources for the fields $\psi(x),\,\bar\psi(x)$ respectively and 
$S_F=(i\hat\partial     -q\hat A(x) -m)^{-1}$ is the Green's function for the Dirac operator. Obviously the
quadratic functional integral in (\ref{Z}) has been done with the rules of the fermionic case.
We simply denote by $Z[A]$ the partition function with vanishing sources: the effective action is then given by $-i$
the times  logarithm of $Z[A]$, namely 
\begin{eqnarray}
\Seff[A]=-i\,\ln Z[A]=-i\,\ln\,\det\,(i\hat\partial-q\hat A(x) -m)=-i\,{\mathrm{tr}}\,\ln\,(i\hat\partial-q\hat A(x)-m)
\label{Seff} 
\end{eqnarray}
In eq. (\ref{Seff}) the determinant and the trace have to be taken both in the space of the coordinates and in the space
of the Dirac variables.

We will show how the FWT can be used for obtaining the explicit form of the effective action, by sketching the
simplest case of a free Dirac particle and of a Dirac particle in a static and uniform magnetic field.
\smallskip

\textbf{(}\textbf{\emph{i}}\textbf{)} \textbf{\emph{The free spinor particle}.} Let first $A^\mu(x)=0$. As
$\gamma^0\equiv\beta={\mathrm{diag}}(+1,+1,-1,-1)$ and therefore
$\det(\gamma^0)=1$, we have to calculate 
\begin{eqnarray}
  \Seff[0]=-i\,\ln\,\det\,\bigl(p_0-(\vec{\alpha}\!\cdot\!\vec{p} +\beta m)\bigr)\,,\qquad
p_0=i\partial/\partial
t\,,~~~\vec{p}=-i\vec{\nabla}\,.
\label{SeffA0_1} 
\end{eqnarray}
If we now recall that the FWT for the free case is generated by 
\begin{eqnarray}
 U_{FW}= \exp\,\Bigl(\, \frac{\arctan(|\vec{p}|/m)}{2|\vec{p}|}\, \beta\,\,\vec{\alpha}\!\cdot\!\vec{p} \,\Bigr)
\label{UFW_Free}
\end{eqnarray}
and we observe that $U_{FW}$ commutes with $p_0$, we can write:
\begin{eqnarray}
\Seff[0]=-i\,\ln\,\det\,\Bigl(p_0-U_{FW}\,(\vec{\alpha}\!\cdot\!\vec{p} +\beta m)\,U_{FW}^{-1}\Bigr)
=-i\,\ln\,\det\,\Bigl(p_0-\beta\,\sqrt{\vec{p}^2+m^2} \, \Bigr)
\label{SeffFA0_2} 
\end{eqnarray}
Due to the diagonal form of $\beta$, the determinant in the Dirac space is immediate and we are reduced to calculating  
\begin{eqnarray}
\Seff[0]=-i\,\ln\,\det\,\Bigl[ \Bigl(i\frac{\partial}{\partial t}
-\sqrt{-\vec{\nabla}^2+m^2}\Bigr)\, \Bigl(i\frac{\partial}{\partial t} +\sqrt{-\vec{\nabla}^2+m^2}\Bigr)   
\Bigr]^2
\label{SeffFA0_3}
\end{eqnarray}
where now the determinant is only the functional determinant in the coordinate space. Hence we obtain the well known
result \cite{IZ,DJ}
\begin{eqnarray}
 \Seff[0]=-i\, {\mathrm{tr}}\,\ln\,\Bigl(\frac{\partial^2}{\partial t^2} -\vec{\nabla}^2+m^2\Bigr)^2
=-2i\, {\mathrm{tr}}\,\ln\,\bigl(\,\square +m^2\,\bigr)\,.
\label{SeffFA0_4}
\end{eqnarray}
The last expression of (\ref{SeffFA0_4}) is calculated, as usual, by inserting complete sets of momentum eigenstates.
Taking into account of the infinite volume factor coming from the scalar product $\bra {\!p\!}\,
p\,\rangle=(2\pi)^4\,\delta^4(0)=\int d^4x\equiv V_4$, we have 
\begin{eqnarray}
 \Seff[0]=-2i\,V_4 \int \frac{d^4p}{(2\pi)^4}\,\ln\,(-p^2+m^2)
\label{SeffFA0_5}
\end{eqnarray}
Finally, recalling that the effective action is the space-time integral of the effective Lagrangian density
$\mathcal{L}_\mathrm{eff}$, we can integrate over $p_0$ and write  the effective potential of our system as
\begin{eqnarray}
 V_{\mathrm{eff}}[0]=-\mathcal{L}_{\mathrm{eff}}[0]=-2\,\int\frac{d^3p}{(2\pi)^3}\,\sqrt{\vec{p}^2+m^2}
=\frac 1{8\pi^2}\int_0^\infty\,\frac{ds}{s^3}\,\exp(-m^2s)
\label{Veff_0}
\end{eqnarray}
where the last expression is a straightforward consequence of the well known representation of the logarithm of an
operator
\begin{eqnarray}
 \ln A = -\int_0^\infty \frac{ds}s\,\exp(-is(A-i\epsilon)\, \quad (\epsilon>0)\,,
\label{ln_int}
\end{eqnarray}
when applied to $A=\square+m^2$.
\smallskip

\textbf{(}\textbf{\emph{ii}}\textbf{)} \textbf{\emph{The spinor particle in a uniform magnetostatic field}.} The next
example we
want to describe is that of a Dirac
particle in a static and uniform magnetic field.
We will assume a vector potential and a corresponding magnetic field given by
\begin{eqnarray}
 A^0=0\,,\qquad \vec{A}=(0,B\,x,0)\,,\qquad \vec{B}=(0,0,B)\,.
\label{vect_A}
\end{eqnarray}
Along the same lines as for the free case, we see that equation (\ref{SeffA0_1}) must now obviously changed into
\begin{eqnarray}
  \Seff[A]=-i\,\ln\,\det\,\bigl(p_0-(\vec{\alpha}\!\cdot\!\vec{\pi} +\beta m)\bigr)\,,\qquad \vec{\pi}=\vec{p}-q\vec{A}
\label{SeffA_1} 
\end{eqnarray}
where $q$ is the particle electric charge. The explicit form of the FWT generator is now more complex, namely 
\cite{BCL4,EK}
\begin{eqnarray}
 U_{FW}=\exp\,\Bigl(\beta\,\mathcal{O}\,\phi\,\Bigr)\,,\qquad \mathcal{O}=\vec{\alpha}\!\cdot\!\vec{\pi},
\qquad\mathcal{O}^2=
{\vec{\pi}^2-q\,\,\vec{\Sigma}\!\cdot\!\vec{B}}
\label{UFW_A}
\end{eqnarray}
where  the angle $\phi$ is given by 
\begin{eqnarray}
 \phi= 1/{(2\sqrt{\mathcal{O}^2})}\,\arctan \Bigl({\sqrt{\mathcal{O}^2}}/{m}\Bigr)
 \label{theta_di_phi}
\end{eqnarray}
By applying the FWT we find now
\begin{eqnarray} 
\Seff[A]=-i\,\ln\,\det\,\Bigl(p_0-U_{FW}\,\mathcal{O}\,U_{FW}^{-1}\Bigr)=-i\,\ln\,\det\,\bigl(p_0-\beta\,\sqrt{\mathcal{
O}^2+m^2})
\label{SeffA_2} 
\end{eqnarray}
Using commutation and anti-commutation relations 
\begin{eqnarray}
[\gamma_5,\vec{\Sigma}]=[\gamma_5,\beta]_+=0\,,
\label{gamma5_Sigma} 
\end{eqnarray}
we have
\begin{eqnarray}
&{}&\!\!\!\!\!\!\!\!\!\!\!\!\!\!\!\!\!\!\!\!\!\!\!\!\!\det\,\bigl(p_0-\beta\,\sqrt{\mathcal{O}^2+m^2})
=\det\,\bigl(\gamma_5(p_0-\beta\,\sqrt{\mathcal{O} ^2+m^2})\gamma_5
)=\det\, \bigl(p_0+\beta\,
\sqrt{\mathcal{O}^2+m^2})=
\spazio{1.0}\cr
&{}&\!\!\!\!\!\!\!\!\!\!\!\!\!\!\!\!\!\!
\sqrt{\det\,\bigl(p_0-\beta\,\sqrt{\mathcal{O}^2+m^2})\,\det\,\bigl(p_0+\beta\,\sqrt{\mathcal{ O}^2+m^2})}
=\sqrt{\det\left(p_0^2-\vec{\pi}^2-m^2+q\,\,\vec{\Sigma}\!\cdot\!\vec{B}\right)}
\label{det_p0_meno_betaphi}
\end{eqnarray}
We can now observe that the matrix operator under the
last square root is in the  diagonal block form
$\,q\,\vec{\Sigma}\!\cdot\!\vec{B}=q\,B\,\mathrm{diag}(\sigma_3,\sigma_3)$,
so that the Dirac space determinant is easily evaluated and the effective action reads
\begin{eqnarray}
 \Seff[A]=-i\,\ln\,\det\,\Bigl(\bigl(-p_0^2+\vec{\pi}^{\,2}+m^2-qB\bigr)\,\bigl(-p_0^2+\vec{\pi}^{\,2}
+m^2+qB\bigr)\Bigr)
\label{SeffA_3} 
\end{eqnarray}
where, again, we are left with the functional determinant on the space-time variables. Substituting the logarithm of
the determinant with the trace of the logarithm we see that we are substantially reduced to the scalar case and we
have to evaluate the quantity
\begin{eqnarray}
 J_\pm=-i\,\mathrm{tr}\,\ln\,(-\pi^2+M_\pm^2)\,,\qquad M_\pm^2=m^2\pm q\,B\,,\qquad \pi^\mu=p^\mu-qA^\mu
\label{J}
\end{eqnarray}
It is well known that there are several ways to calculate the above trace. We can use the knowledge of the
eigenfunctions of the operator $\,-\pi^2+M_\pm^2\,$, or we can solve the Heisenberg equation of motion and find the
evolution operator in the Schwinger proper time formalism, or else we can calculate the evolution operator by
means of a path integral \cite{BBC2001}, which is possible as we have reduced the problem in a quadratic Bose form. We
notice that for a constant magnetic field we could have proceeded also directly by integrating over the odd variables;
since this procedure cannot be directly exported in the presence of an anomalous magnetic moment, we show how the other
methods, which generally apply, will be used.

Specifying equation (\ref{ln_int}) to our present case, we write
\begin{eqnarray}
J_\pm=i\,\mathrm{tr}\!\int_0^\infty \frac{ds}{s}\,\exp\Bigl(-is\,(-\pi^2+M_\pm^2-i\epsilon)\Bigr)
=-i\,\mathrm{tr}\!\int_{-\infty}^0 \frac{ds}{s}\,\exp\Bigl(-is\,(\pi^2-M_\pm^2+i\epsilon)\Bigr)
\label{JB}
\end{eqnarray}
The explicit form of the wave operator that appears in (\ref{JB}) is
\begin{eqnarray}
-(\pi^2-M_\pm^2)=\frac{\partial^2}{\partial t^2}-\vec{\nabla}^2+2\,i\,qB\,x\,\frac{\partial}{\partial y}+ q^2B^2x^2 +
M_\pm^2
\label{waveop_B}
\end{eqnarray}
and we denote its eigenfunctions by
$\ket{p_0,p_y,p_z;n}\equiv\ket{p_0,p_y,p_z}\ket{\varphi_n}$.
The first term in the right hand side is the factor corresponding to the conserved momenta $p_0,\,p_y,\,p_z$,
while $\varphi_n(x)=\langle x\,|\,\varphi_n\rangle$ solves the eigenvalue problem
\begin{eqnarray}
\Bigl[\,-\frac{\partial^2}{\partial
x^2}+q^2B^2\Bigl(x-\frac{p_y}{qB}\Bigr)^2\,\Bigr]\varphi_n(x)=\lambda_n\,\varphi_n(x)
\label{LandauLevels}
\end{eqnarray}
with $\lambda_n=(2n+1)\,qB$, being the eigenvalues for a harmonic oscillator with frequency $qB$. We assume the
functions $\varphi_n(x)$ normalized to unity.
Inserting the appropriate completeness relations the conserved momenta give rise to an
infinite factor $(2\pi)^3\delta^3(0)=L_2L_3L_0$, where $L_2\,, L_3\,$ refer to the space directions $y$ and $z$
and $L_0$ to time. After some lengthy but straightforward 
calculations, the
quantity $J$ of equation (\ref{JB}) becomes 
\begin{eqnarray}
\!\!\!\!\!\!\!\!J_\pm\!\!\!\!&=&\!\!\!\!i\,{L_2L_3L_0}\,\int_0^\infty\frac{ds}{s}\,\int_{-\infty}^{\infty} 
\frac{dp_0}{2\pi}\,\int_{-\infty}^{\infty} 
\frac{dp_z}{2\pi}\,\int_0^{qBL_1} \frac{dp_y}{2\pi}\,\spazio{1.0}\cr
\!\!\!\!\!\!\!\!\!\!\!\!&{}&\!\!\!\! \qquad \qquad\qquad\qquad \,\sum_{n=0}^{\infty}
\,\exp\Bigl[-is\,\Bigl(-p_0^2+p_z^2+M_\pm^2+(2n+1)\,qB-i\epsilon\Bigr)\Bigr]\spazio{1.0}\cr
\!\!\!\!\!\!\!\!\!\!\!\!&=&\!\!\!\!i\,\frac{L_1L_2L_3L_0}{8\pi^2}\int_0^\infty \frac{ds}{s^3}\,(s\,q B)\,\exp(\mp\,
is\,q B)\,\sum_{n=0}^{\infty}
\exp\Bigl(-is\,\bigl((2n+1)\,q B+m^2-i\epsilon\bigr)\Bigr)
\label{JB_1}
\end{eqnarray}
where the integration in $p_y$ reconstructs the four dimensional volume $V_4$ and the contribution of the $p_0$ and
$p_z$ integrations is, respectively, $\exp(\pm\,i\pi/4)\,(\pi/s)^{1/2}$.
In order to find the effective action we must add the contributions of $M_+$ and $M_-$: after summing the series we find
\begin{eqnarray}
\Seff[A]=J_+ +J_-=\frac{L_1L_2L_3L_0}{8\pi^2}\int_0^\infty \frac{ds}{s^3}\,\exp\bigl(-is(m^2-i\epsilon)\bigr)\,(s\,q B)\,\cot(s\,q B)
\label{SeffA_4}
\end{eqnarray}
Some care is in order for the calculation of this integral. In the first place because of the poles of $\cot(s\,q
B)$ on the real axis. Secondly, because of possible counterterms that should be added in (\ref{SeffA_4}) in order to
renormalize the effective Lagrangian $\mathcal{L}_{\mathrm{eff}}[A]=\Seff[A]/V_4$. The problem of the poles of the
integrand on the real axis is solved by observing that the integral converges for $\mathrm{Im}(s)<0$, so that the
integration path can be deformed from the positive real semi-axis to the negative imaginary semi-axis, $s\mapsto -is$.
We can next normalize to unity the partition function for a vanishing field. This amounts to considering $Z[A]/Z[0]$ 
instead of $Z[A]$. On the effective action this normalization has the effect of subtracting the free part from the
result in the presence of a magnetic field, so that $\Seff[A=0]=0$. After these two steps we get the effective
Lagrangian
\begin{eqnarray}
\mathcal{L}_\mathrm{eff}[A]=-\frac 1{8\pi^2}\int_0^\infty \frac{ds}{s^3}\,\exp(-sm^2)\,(s\,q B)\,(\coth(s\,q B)-1)
\label{SeffA_5}
\end{eqnarray}
We easily see that the integral in (\ref{SeffA_5}) is still divergent at $s=0$ and needs renormalization. We thus
assume the charge $q$ and the magnetic field $B$ of (\ref{SeffA_5}) as bare quantities $q_0,B_0$; we consider
the expansion $(s\,q_0 B_0)\,(\coth(s\,q_0 B_0)-1)=(1/3)\,(s\,q_0 B_0)^2+O(s^4)$; we add to
$\mathcal{L}_\mathrm{eff}[A]$ the contribution of the tree level, namely $-(1/2)B_0^2$; we rewrite (\ref{SeffA_5})
by adding and subtracting the divergent part. We finally obtain
\begin{eqnarray}
 \mathcal{L}_\mathrm{eff}[A]=\!\!\!\!&-&\!\!\!\!\frac 12\,B_0^2\,\Bigl( 1+\frac 1{12\pi^2}\,q_0^2\int_0^\infty
\frac{ds}{s}\exp(-sm^2) \Bigr)\spazio{1.2}\cr 
\!\!\!\!&-&\!\!\!\!\frac 1{8\pi^2}\int_0^\infty \frac{ds}{s^3}\,\exp(-sm^2)\,(s\,q_0
B_0)\,\Bigl(\coth(s\,q_0 B_0)-\frac 13\,(s\,q_0 B_0)^2-1\Bigr)
\label{SeffA_6}
\end{eqnarray}
We now introduce the renormalization constant for the wave function, the renormalized charge and
magnetic field:
\begin{eqnarray}
Z_3^{-1}=  1+\frac 1{12\pi^2}\,q_0^2\int_0^\infty
\frac{ds}{s}\exp(-sm^2),\qquad q=q_0\,Z_3^{\frac 12},\qquad B=B_0\,Z_3^{-\frac 12}.
\label{Z3}
\end{eqnarray}
We have $q_0 B_0=q B$ and we can write the final expression of the renormalized effective action \cite{Don} 
\begin{eqnarray}
\mathcal{L}_\mathrm{eff}[A]=-\frac12 B^2-\frac 1{8\pi^2}\int_0^\infty \frac{ds}{s^3}\,\exp(-sm^2)\,(s\,q
B)\,\Bigl(\coth(s\,q B)-\frac 13\,(s\,q B)^2-1\Bigr)
\label{SeffA_8}
\end{eqnarray}
\medskip

\sect{Effective action for neutral fermions with a.m.m. in static magnetic and electric fields.}

Let us recall the Hamiltonian form of the Dirac equation for a fermion with a.m.m. $\mu$ interacting
with
an external electromagnetic field:
\begin{eqnarray}
 i\,\frac{\partial\psi(t,\vec{x})}{\partial t} = H_{\mathrm{D}}\psi(t,\vec{x})\,,\qquad
H_{\mathrm{D}}=\vec{\alpha}\!\cdot\!\vec{\pi}+qA_0+\beta
m+\frac{e\mu}{8m}\,\sigma_{\rho\nu}\,F^{\rho\nu}\,,
\label{AnomaloDirac}
\end{eqnarray}
where 
\begin{eqnarray}
\sigma^{\rho\nu}=\frac i2\,[\gamma^{\rho},\gamma^{\nu}]\,,\qquad 
F^{\rho\nu}(x)=\frac{\partial A^{\nu}}{\partial x_{\rho}}-\frac{\partial A^{\rho}}{\partial x_{\nu}}\,,
\label{sigma_F}
\end{eqnarray}
and, as usual, $\vec{\pi}=\vec{p}-q\vec{A}$. 

In addition to (\ref{AnomaloDirac}) we can also use an equivalent expression for the wave equation, obtained by means
of the Pauli-Gursey unitary transformation generated by applying $\,\exp\bigl[i(\pi/4)\gamma_5\bigr]$ to the spinor
wave function \cite{PaGu,BBC2001,BGPJP} and, correspondingly, by mapping the Hamiltonian $H_\mathrm{D}$ into the
Hamiltonian 
$H_\mathrm{PG}=\,\exp\bigl[i(\pi/4)\gamma_5\bigr]\,H_\mathrm{D}\,\,\exp\bigl[-i(\pi/4)\gamma_5\bigr]$. By substituting
$F^{\rho\nu}$ by its explicit expression in terms of the electric and the magnetic field, we can write $H_\mathrm{D}$
and $H_\mathrm{PG}$ as
\begin{eqnarray}
 H_{\mathrm{D}}\!\!\!&=&\!\!\!\vec{\alpha}\!\cdot\!\vec{\pi}+qA_0+\beta\,m +\frac{ie\mu}{4m}\,\beta\,\vec{\alpha}
\!\cdot\!\vec{E}-\frac{e\mu}{4m}\,\beta\,\vec{\Sigma}\!\cdot\!\vec{B}\spazio{1.2} \cr
 H_{\mathrm{PG}}\!\!\!&=&\!\!\!\vec{\alpha}\!\cdot\!\vec{\pi}+qA_0-i \beta\gamma_5\, m
+\frac{e\mu}{4m}\,\beta\gamma_5\,\vec{\alpha}\!\cdot\!\vec{E}
+\frac{ie\mu}{4m}\,\beta\gamma_5\,\vec{\Sigma}\!\cdot\!\vec{B}
\label{HD_HPG}
\end{eqnarray}
where $\vec{\Sigma}=\gamma_5\vec{\alpha}$. The effective action can be equivalently obtained by the Dirac or the
Pauli-Gursey representation and reads
\begin{eqnarray}
\Seff[A]=-i\,\ln\,\det\Bigl(i\frac{\partial}{\partial
t}-H_{\mathrm{D}}\Bigr)=-i\,\ln\,\det\Bigl(i\frac{\partial}{\partial t}-H_{\mathrm{PG}}\Bigr)
\label{SeffA_7}
\end{eqnarray}
where $H_{\mathrm{D}}$ and $H_{\mathrm{PG}}$ are given in (\ref{HD_HPG}).

Since we want to calculate the effective action by using the FWT along the lines described in Section 2, we
find it useful to recall some results
determined in \cite{BGPJP} and related to the cases in which the FWT can be found in closed form, namely the cases of a 
neutral fermion with a.m.m. in a static and uniform magnetic field or in a static and uniform
electric field. Letting first $q=0$ and $\vec{E}=0$ in the Hamiltonian $H_{\mathrm{PG}}$ in (\ref{HD_HPG}) and
applying to it a unitary transformation generated by
\begin{eqnarray}
U=\exp\bigl(\beta\mathcal{O}\phi\bigr)\,,\qquad \mathcal{O}=H_{\mathrm{PG}}\,,\qquad
\phi=\frac{\pi}{4}\,\bigl(\mathcal{O}^{\,2}\bigr)^{-\frac 12} 
\label{FW_q0_E0}
\end{eqnarray}
we obtain a transformed even Hamiltonian of the form
\begin{eqnarray}
 \widetilde{H}_{\mathrm{PG}}=UH_{PG}U^{-1}=\beta\,\Bigl[\,\vec{p}^{\,\,2}+m^2+\Bigl(\frac{e\mu}{4m}\Bigr)^2\vec{B}^{\,2}
-\frac { e\mu } { 2
} \,
\vec { \Sigma
}\!\cdot\!\vec{B}-\frac{e\mu}{4m}\,\beta\,\vec{\Sigma}\!\cdot\!\bigl(\vec{B}\times\vec{p}-\vec{p}\times\vec{B}\bigr)\,
\Bigr]^{\frac12} 
\label{FW_HPG}
\end{eqnarray}
Since $(\vec{\Sigma}\!\cdot\!\vec{B})^2=B^2\,I$, letting $\vec{p}=0$ in (\ref{FW_HPG}), we immediately see that
the threshold levels are given by $\,E_0=\pm\bigl(m\pm (e\mu/4m)B\bigr)$ with an energy gap between the intermediate
levels
\begin{eqnarray}
\Delta E_0=2\Bigl(m-\frac{|e\mu|}{4m}B\Bigr)
\label{DeltaE0B}
\end{eqnarray}
Introducing the momentum components longitudinal and transverse to the magnetic field $\vec{B}$,
\begin{eqnarray}
\vec{p}_\ell={(\vec{p}\!\cdot\!\vec{B})\vec{B}}\,/\,{\vec{B}^{\,2}}\,,\qquad  \vec{p}_t=\vec{p}-\vec{p}_\ell
\label{p-ellt-B}
\end{eqnarray}
from the dispersion relation (\ref{FW_HPG}) we can write the energy eigenvalues
\begin{eqnarray}
p_0=E=\pm\,\Bigl[\,p_\ell^2+\Bigl((p_t^2+m^2)^{\frac 12}\pm \frac{e\mu B}{4m}\Bigl)^2\,\Bigl]^{\frac 12}
\label{E-eigen-B}
\end{eqnarray}

In a completely analogous way, if in (\ref{HD_HPG}) we consider $H_{\mathrm{D}}$ with $q=0$, $\vec{B}=0$ and we make a
unitary
transformation generated by
\begin{eqnarray}
U=\exp\bigl(\beta\mathcal{O}\phi\bigr)\,,\qquad
\mathcal{O}=\vec{\alpha}\!\cdot\!\Bigl(\vec{p}-\frac{ie\mu}{4m}\beta\vec{E}\Bigr)\,,\qquad
\phi=\frac1{2\sqrt{\mathcal{O}^2}}\,\arctan\Bigl(\frac{\sqrt{\mathcal{O}^2}}m\,\Bigr) 
\label{FW_q0_B0}
\end{eqnarray}
we obtain a transformed even Hamiltonian of the form
\begin{eqnarray}
\!\!\!
\widetilde{H}_{\mathrm{D}}\!=\!UH_{D}U^{-1}=\beta\,\Bigl[\,\vec{p}^{\,\,2}+m^2+\Bigl(\frac{e\mu}{4m}\Bigr)^2\vec{E}^{\,2
} -\frac { e\mu } { 4m
} \, \beta\,\vec { \nabla
}\!\cdot\!\vec{E}-\frac{e\mu}{4m}\,\beta\,\vec{\Sigma}\!\cdot\!\bigl(\vec{E}\times\vec{p}-\vec{p}\times\vec{E}\bigr)\,
\Bigr]^{\frac 12} 
\label{FW_HD}
\end{eqnarray}
The threshold energy levels are now $\,E_0=\pm\bigl(m^2+(e\mu/4m) E)^2\bigr)^{\frac 12}$ with energy gap
\begin{eqnarray}
\Delta E_0=2\Bigl(m^2+\bigl(\frac{e\mu}{4m}\bigr)^2 \,E^2\Bigr)^{\frac 12}
\label{DeltaE0E}
\end{eqnarray}
and energy eigenvalues
\begin{eqnarray}
p_0=E=\pm\,\Bigl[\,p_\ell^2+m^2+\Bigl(p_t\pm\frac{e\mu}{4m}\,E\Bigr)^2\,\Bigr]^{\frac 12}
\label{E-eigen-E}
\end{eqnarray}
where now $\vec{p}_\ell$ and $\vec{p}_t$ are defined by substituting $\vec{E}$ to $\vec{B}$
in (\ref{p-ellt-B}).  

Let us examine in more details the two cases.
\smallskip

\textbf{(}\textbf{\emph{i}}\textbf{)} \textbf{\emph{The spinor with a.m.m. in a uniform magnetostatic field.}} We want
to calculate the functional determinant
(\ref{SeffA_7}). In the momentum representation, after the appropriate FWT the determinant reads
\begin{eqnarray}
I=\det\Bigl(p_0-\widetilde{H}_{\mathrm{PG}}\Bigr)
\end{eqnarray}
where $\widetilde{H}_{\mathrm{PG}}$ is given in (\ref{FW_HPG}).

We find it convenient to introduce the following notation:
\begin{eqnarray}
Q_1=p_0\,,\quad Q_2=\vec{p}^{\,\,2}+m^2+\Bigl(\frac{e\mu\vec{B}}{4m}\Bigr)^2\,,\quad
Q_3=\frac{e\mu}{2}\,\vec{\Sigma}\!\cdot\!\vec{B}+\frac{e\mu}{2m}\,\beta\,\vec{\Sigma}\!\cdot\!(\vec{B}\times\vec{p})\,,
\label{V123}
\end{eqnarray}
so that we want to find $\det Q$ where
\begin{eqnarray}
Q=Q_1-\beta\,\sqrt{Q_2-Q_3}\,\quad\mathrm{ with}\quad \bigl[V_i,V_j\bigr]=0\,,\quad i,j=1,2,3\,.
\label{V_e_comm}
\end{eqnarray}
We therefore need, in the first place, the eigenvalues of $V$ in the spinor space.
Observing that
\begin{eqnarray}
Q_3^2=\Bigl(\frac{e\mu}{2}\Bigr)^2\,\Bigl[\vec{B}^{\,2} +\frac
1{m^2}\,\Bigl(\vec{B}^{\,2}\,\vec{p}^{\,\,2}-(\vec{p}\,\cdot\,\vec{B})^2\Bigr)\Bigr]\equiv \lambda_{3B}\,I
\label{V3B}
\end{eqnarray}
we see that the eigenvalues of $Q_3$ are $\pm\sqrt{\lambda_{3B}}$ each one with multiplicity two. 
We then follow the same procedure we used in the previous section and recalling the relations
(\ref{gamma5_Sigma}) we write 
\begin{eqnarray}
(\det
Q)^2\!=\!\det\Bigl[\Bigl(Q_1\!-\!\beta\sqrt{Q_2\!-\!Q_3}\Bigr)\Bigl(\gamma_5(Q_1-\beta\sqrt{Q_2\!-\!Q_3}
)\gamma_5\Bigr)\Bigr ]
\!=\!\det\Bigl(Q_1^2\!-\!(Q_2\!-\!Q_3)\Bigr)\,
\label{detV2}
\end{eqnarray}
Therefore, taking into account the multiplicity of the eigenvalues, we have that the determinant in the spinor space
reads
\begin{eqnarray}
\det Q\!=\!\prod_{\eta=\pm 1
}\,\Bigl[\,p_0^2-\vec{p}^{\,\,2}-m^2-\Bigl(\frac{e\mu\vec{B}}{4m}\Bigr)^2-\eta\Bigl(\frac{e\mu} 2\Bigr)\,
\sqrt{\vec{B}^{\,2}+\frac 1{m^2}\bigl(\vec{B}^{\,2}\vec{p}^{\,\,2}-(\vec{p}\!\cdot\!\vec{B})^2} \,
\Bigr]
\label{detV_final}
\end{eqnarray}
Introducing now the momentum components longitudinal and transverse to the magnetic field $\vec{B}$,
\begin{eqnarray}
\vec{p}_\ell={(\vec{p}\!\cdot\!\vec{B})\vec{B}}\,/\,{\vec{B}^{\,2}}\,,\qquad  \vec{p}_t=\vec{p}-\vec{p}_\ell
\label{elltB}
\end{eqnarray}
the effective action reduces to the following form
\begin{eqnarray}
&{}&\Seff[A]=-i\,\ln\det\Bigl[\,\prod_{\eta=\pm1}\Bigl(p_0^2-\vec{p}_\ell^{\,\,2}-\,
\Bigl(\sqrt{
\vec { p } _t^ { \, \, 2 } +m^2 }
+\eta\,\frac{ e\mu B}{4m}\Bigr)^2\,\Bigr)\,\Bigl]\,.
\end{eqnarray}
By means of the usual identity $\,\mathrm{ln}\, \mathrm{det}= \mathrm{tr} \,\mathrm{ln}\,$ and by the
representation (\ref{ln_int}) for the logarithm, we can write 
\begin{eqnarray}
&{}&\!\!\!\!\!\!\!\!\!\!\!\!\!\!\!\!\!\!\!\!\!\!\!\Seff[A]=\cr
&{}&\!\!\!\!\!\!\!\!\!\!\!\!\!\!\!\!{iV_4}\sum_{\eta=\pm1}\int\frac{d^4p}{(2\pi)^4}\int_0^\infty\frac{ds}{s}
\,\exp\Bigl[-is\Bigl(-p_0^2+p_\ell^2+\Bigl((p_t^2+m^2)^{\frac 12}+\eta(e\mu/4m)\,B\Bigl)^2-i\epsilon\Bigr)\Bigr]
\label{seff_a_mag}
\end{eqnarray}
Calculating the integrals over the zero
and parallel components of the momentum, making the complex rotation $s\rightarrow -is$ and dividing by the four
dimensional infinite volume $V_4$, we find an effective
Lagrangian
\begin{eqnarray}
\mathcal{L}_\mathrm{eff}=-\frac 1{8\pi^2}\sum_{\eta=\pm1}\int_0^\infty\frac{ds}{s^2}\int_0^\infty dp_t\,p_t
\,\exp\Bigl[\,-s\Bigl((p_t^2+m^2)^{\frac 12}+\eta(e\mu/4m)\,B\Bigr)^2\,\Bigr]
\label{leff_1}
\end{eqnarray}
Let us now define 
\begin{eqnarray}
\gb=(|e\mu|\,B)/(4m^2),\qquad\qquad  z=m^2s,
\label{pargammaB}
\end{eqnarray}
and, accordingly, change the variables in (\ref{leff_1}).  We then normalize the effective Lagrangian by
subtracting the contribution for a vanishing magnetic field, so that $\mathcal{L}_\mathrm{eff}[B=0]=0$
and we add the free contribution of the magnetic field. 
After some lengthy but straightforward calculations we then have 
\begin{eqnarray}
&{}&\!\!\!\!\!\!\!\!\!\!\!\!\!\!\!\!\!\!\!\!\!\mathcal{L}_\mathrm{eff}=
-\frac12{B^2} -\frac
{m^4}{16\pi^2}\int_0^\infty\frac{dz}{z^3}\,\Bigl[
\exp\bigl(-(\gb+1)^2z\bigr)+\exp\bigl(-(\gb-1)^2z\bigr)-2\exp\bigl(-z\bigr)\Bigr]
\cr\spazio{1.8}
&{}&\!\!\!\!\!\!\!\!\!\!\!\!\!\!\!\!\!\!-\frac{m^4\gb}{16\pi^2}\int_0^\infty\frac{dz}{z^2}\,\int_{-1}^{1}du\,
\Bigl[(\gb+1)\exp\bigl(-(\gb+1)^2u^2z\bigr)+(\gb-1)\exp\bigl(-(\gb-1)^2u^2z\bigr)\Bigr]
%&{}&
\label{lagB_norm}
\end{eqnarray}

We finally consider the renormalization of the effective Lagrangian. Indeed the expression (\ref{lagB_norm}) is
still singular for $z=0$ and we will rewrite it by subtracting the leading divergent terms
obtaining a minimally regularized effective Lagrangian
\begin{eqnarray}
\mathcal{L}_\mathrm{eff}\!\!\!&=&\!\!\!-\frac12 B^2-\frac{m^4}{16\pi^2}\int_0^\infty\frac{dz}{z^3}\Bigl[\,
\exp{(- \bigl( \gb+1 \bigr) ^{2}z)}+  \exp{(- \bigl( \gb-1 \bigr)
^{2}z)} -2\,\exp{(-z)}
\cr\spazio{1.4}
&{}& +\,\gb z\int _{-1}^{1}{du}\,\,\bigl[\, \bigl( \gb+1 \bigr) {\exp{(- \bigl( \gb+1 \bigr) ^{2}{u}^{2}z)}}+ \bigl(
\gb-1 \bigr)
\exp{(- \bigl( \gb-1 \bigr) ^{2}{u}^{2}z)}\,\bigr]
\cr\spazio{1.4}
&{}& +\exp(-z) \Bigl(-\,2\,{\gb}^{2}z-\,{\gb}^{2} \Bigl( 4-\frac13{\gb}^{2}\Bigr) {z}^{2}\,\Bigr)\Bigr]
+ f.t. 
\label{lagB_reg}
\end{eqnarray}
 where $f.t.$ indicates the presence of possible finite terms to be determined by a specific renormalization
prescription.

%============================================================================================================
\bigskip
\textbf{(}\textbf{\emph{ii}}\textbf{)} \textbf{\emph{The spinor with a.m.m. in an electrostatic field.}} We want now to
calculate the determinant
\begin{eqnarray}
I=\det\Bigl(p_0-\widetilde{H}_{\mathrm{D}}\Bigr)
\end{eqnarray}
with $\widetilde{H}_{\mathrm{D}}$ given in (\ref{FW_HD}). Since we are considering uniform fields we have the relations
\begin{eqnarray}
\vec{\nabla}\!\cdot\!\vec{E}=0\,,\qquad \vec{E}\times\vec{p}=-\vec{p}\times\vec{E}\,,
\end{eqnarray}
so that, letting
\begin{eqnarray}
Q_1=p_0\,,\quad Q_2=\vec{p}^{\,\,2}+m^2+\Bigl(\frac{e\mu\vec{E}}{4m}\Bigr)^2\,,\quad
Q_3=\frac{e\mu}{2m}\,\beta\,\vec{\Sigma}\!\cdot\!(\vec{E}\times\vec{p})\,,
\label{VE123}
\end{eqnarray}
we have to evaluate $\det Q$ where $Q$ is again given by (\ref{V_e_comm}) with the definitions  
(\ref{VE123}).
We now see that
\begin{eqnarray}
Q_3^2=\Bigl(\frac{e\mu}{2m}\Bigr)^2\,\Bigl(\vec{E}^{\,2}\,\vec{p}^{\,\,2}-(\vec{p}\,\cdot\,\vec{E})^2\Bigr) \equiv
\lambda_{3E}\,I
\label{V3E}
\end{eqnarray}
and therefore the eigenvalues of $Q_3$ are $\pm\sqrt{\lambda_{3E}}$ each one with multiplicity two.
Making again the same calculations of the previous paragraph and introducing the momentum components longitudinal and
transverse to the electric field, given by (\ref{elltB}) with $\vec{B}$ substituted by $\vec{E}$, we finally arrive to
the effective action
\begin{eqnarray}
\Seff[A]=-i\ln\det\Bigl[\,\prod_{\eta=\pm1}
\Bigl(p_0^2-{p}_\ell^{\,\,2}-m^2-\Bigl({p}_t
+\eta\frac{ e\mu}{4m}{E}\Bigr)^2\,\Bigr)\Bigr]\,.
\label{Seff_es}
\end{eqnarray}
The effective Lagrangian, therefore, can
be written as
\begin{eqnarray}
&{}&
\!\!\!\!\!\mathcal{L}_\mathrm{eff}=\frac{i}{8\pi^3}\sum_{\eta=\pm1}\int_0^{\infty}\frac{ds}{s}
\int_{-\infty}^{\infty}dp_0
\int_{-\infty}^{\infty}dp_\ell\int_0^{\infty}p_t\,dp_t
\cr
&{}& 
\qquad\qquad\qquad\qquad\qquad\cdot\,\exp\Bigl[-is\Bigr( -p_0^2+p_\ell^2+m^2+\Bigl(p_t+\eta\frac{e\mu
E}{4m}\Bigr)^2-i\epsilon
\Bigr)\Bigr]
\label{Leff_es} 
\end{eqnarray}
Introducing
\begin{eqnarray}
\ge=(|e\mu|\,E)/(4m^2),\qquad \qquad z=m^2s,
\label{pargammaE}
\end{eqnarray}
and going through steps analogous to those of item $(i)$, we arrive to the normalized effective Lagrangian
which vanishes for a vanishing electric field:
\begin{eqnarray}
\!\!\!\!\!\!\!\!\!\!\mathcal{L}_\mathrm{eff}\!\!\!&=&\!\!\!
\frac12{E^2} -\frac
{m^4}{8\pi^2}\int_0^\infty\frac{dz}{z^3}\,\exp\bigl(-z\bigr)\Bigl[\exp\bigl(-\ge^2z\bigr)-1\Bigr]
\cr\spazio{1.8}
&{}&\!\!\!\phantom{XXXXXXX}
-\frac
{m^4\ge^2}{8\pi^2}\int_0^\infty\frac{dz}{z^2}\,\exp\bigl(-z\bigr) \int_{-1}^{1}du\, \exp\bigl(-\ge^2u^2z\bigr)
\label{lagE_norm}
\end{eqnarray}
The expression (\ref{lagE_norm}) is again singular for $z=0$. We can subtract the leading divergent
terms obtaining a minimally regularized effective Lagrangian
\begin{eqnarray}
\!\!\!\!\!\!\!\!\!\!\!\!\!\!\!\mathcal{L}_\mathrm{eff}\!\!\!&=&\!\!\!\frac12{E^2}- \frac
{m^4}{8\pi^2}\int_0^\infty\frac{dz}{z^3}\,\exp\bigl(-z\bigr)\,\cdot
\cr\spazio{1.0}
&{}& \Bigl[\, \exp\bigl(-\ge^2z\bigr)-1+
2\ge^2z\int_0^1 du\,\bigl[\,\exp\bigl(-\ge^2u^2z\bigr)\,\bigr]-\ge^2z+\frac 16 \ge^4z^2\,\Bigr] + f.t. 
\label{lagE_reg}
\end{eqnarray}
 where $f.t.$ indicates the presence of possible finite terms to be determined by a specific renormalization
prescription.

\sect{Finite temperature extension.}

In this final section we will extend our analysis to finite temperatures: we shall therefore calculate
the partition functions for the fermion systems studied in the previous section. 
%and eventually we shall draw some conclusions.

We recall that  the finite temperature extension of the theory is done for the stationary case and for
uniform fields we can
substitute $V_4$ with $\beta V$, where $V=L_1L_2L_3$ is the spatial volume and $\beta=1/T$ in units in
which the Boltzmann constant $k$  is taken unity. The energy is then discretized to the Matsubara frequencies
$p_0=(i\pi/\beta)(2j+1)$ and the integral $\int dp_0/(2\pi)$ is therefore changed into 
$(i/\beta)\,\sum_{j=-\infty}^{+\infty}$ \cite{DJ}. The sum is easily dealt with by using the Poisson relation
that, for the fermionic case \cite{BCP}, reads
\begin{eqnarray}
\sum_{j=-\infty}^{+\infty} \, \delta\Bigl(p_0-\frac\pi\beta \,(2j+1)\Bigr)=
\beta\,\sum_{j=-\infty}^{+\infty} (-1)^j \exp({ij\beta p_0})\,.
\label{Poisson}
\end{eqnarray}  
In order to show the subsequent steps of the calculation, take first a Dirac particle in a magnetic field whose the
effective action is
$\Seff[A]=J_++J_-$ with $J_\pm$ given in (\ref{JB_1})\,. 
We have
\begin{eqnarray}
&{}& \!\!\!\!\!\!\!\mathcal{L}_\mathrm{eff}=i\,\int_0^{\infty}\frac{ds}{s}\,
\sum_{\pm,n=0}^{\infty}\, \exp(-is M_\pm^2)\,i
\sum_j (-1)^j \int_{-\infty}^{+\infty}\frac{dp_0}{2\pi}
\int_{-\infty}^{+\infty}\frac{dp_z}{2\pi}\spazio{1.0}\cr
&{}&\qquad \exp(i(-p_0^2-p_z^2)s)\,\exp(ij\beta p_0)\,\frac{qB}{2\pi\,}\exp(-iqB(2n+1)s)
\end{eqnarray}
Integrating over $p_0$ and $p_z$, summing over $\pm$ and $n$ and finally rotating the integration path of the
variable $s$ as in the zero temperature case, we get 
\begin{eqnarray}
\mathcal{L}_\mathrm{eff}=
\-\frac{1}{8\pi^2} \,\int_{0}^{+\infty}\,\frac{ds}{s^3}\,\-\exp{(-m^2s)}\,(sqB)\,\coth(qBs) 
\sum_{j=-\infty}^{+\infty} (-1)^{j+1} \exp\Bigl({\frac{-j^2\beta^2}{4s}}\Bigr)
\label{Seff_BT_1}
\end{eqnarray}
Since the free Helmholtz energy density $\mathit{f}$ is defined as 
\begin{eqnarray}
\mathit{f}=\frac FV= -\frac{\ln Z}{\beta V} =-\mathcal{L}_{\mathrm{eff}}\,,
\label{z}
\end{eqnarray}
from (\ref{z}) we clearly see that the zero temperature contribution is simply obtained by taking $j=0$, so that we can
write
$\mathit{f}$ as
\begin{eqnarray}
&{}& \!\!\!\!\!\!\!\!\!\!\!\!\!\!\!\!\!\!\!\!\!\!\!\!\!\!\!\mathit{f}=\frac 1{8\pi^2}\int_0^\infty
\frac{ds}{s^3}\,\exp(-sm^2)\,(s\,q
B)\,\Bigl(\coth(s\,q B)-\frac 13\,(s\,q B)^2-1\Bigr)+\spazio{1.0}\cr
&{}& \frac 1{8\pi^2}\int_0^\infty \frac{ds}{s^3}\,\exp(-sm^2)\,(s\,q
B)\,\coth(s\,q B)\,{ {{\sum}_{{j\not= 0} }}^{}}(-1)^j \exp{\Bigl(\frac{-j^2\beta^2}{4s}\Bigr)}
\label{z_T}
\end{eqnarray}
where the zero temperature part is the one we have previously studied, neglecting the tree level contribution
$-(1/2)B^2$.
In the temperature dependent term we make the expansion
\begin{eqnarray}
x\,\coth(x) = x\,\Bigl[ \,\sum_{n=0}^{\infty} ( \exp(-2nx)+\sum_{n=0}^{\infty} ( \exp(-2(n+1)x) \Bigr]
\end{eqnarray}
and we recall the integral representation for the $K$ Bessel functions \cite{MO}:
\begin{eqnarray}
 K_\nu(2(z\zeta)^{\frac 12})=\frac 12 \Bigl(\frac \zeta z\Bigr)^{\nu/2}\,\int_0^\infty dt \,\exp\Bigl(-zt-\frac\zeta
t\Bigr)\,t^{-\nu-1}
\label{Knu}
\end{eqnarray}
We then see that in (\ref{z_T}) we are left with a series in terms of the Bessel function $K_1$.
By means of the further integral representation
\begin{eqnarray}
 K_1(z)=\int_0^\infty dt \,\exp(-z\sqrt{1+t^2})
\label{K1}
\end{eqnarray}
we get
\begin{eqnarray}
 \sum_{j=1}^{\infty}\frac {(-1)^j}j \, u\,K_1(j\beta u) = -\int_0^\infty dt \,
\ln\Bigl(1+\exp(-\beta\sqrt{t^2+u^2})\Bigr)\,,
\label{quattronove}
\end{eqnarray}
and we finally obtain the following expression for the free Helmholtz energy density:
\begin{eqnarray}
&{}& \!\!\!\!\!\!\!\!\!\!\!\!\!\!\!\!\!\!\!\!\!\!\!\!\!\!\!\mathit{f}=\frac 1{8\pi^2}\int_0^\infty
\frac{ds}{s^3}\,\exp(-sm^2)\,(s\,q
B)\,\Bigl(\coth(s\,q B)-\frac 13\,(s\,q B)^2-1\Bigr)+\spazio{1.0}\cr
&{}& \qquad -\frac{2qBT}{\pi}\,\sum_{n=0}^{\infty}\int_0^\infty\frac{dp_z}{2\pi}\Bigl[\,\ln\Bigl(1+\exp(-\beta\sqrt{p_z^2+m^2+2nqB}\,)\Bigr)\spazio{1.0}\cr
&{}&\qquad\qquad\qquad\qquad~~ +\ln\Bigl(1+\exp(-\beta\sqrt{p_z^2+m^2+2(n+1)qB}\,)\Bigr)
\Bigr]
\end{eqnarray}

We will now apply these ideas to the fermions with anomalous magnetic moment we have treated in Section 3 in order to
calculate the close form for their free energy density.

\medskip
\textbf{(}\textbf{\emph{i}}\textbf{)} \textbf{\emph{The free energy for a spinor with a.m.m. in a magnetostatic
uniform field.}}
Starting from (\ref{seff_a_mag}) and following the steps previously outlined, a simple calculation leads to the
following expression for the effective Lagrangian:
\begin{eqnarray}
\mathcal{L}_\mathrm{eff}= \mathcal{L}_\mathrm{eff}(B,0)+\mathcal{L}_\mathrm{eff}(B,T)
\end{eqnarray}
where $\mathcal{L}_\mathrm{eff}(B,0)$ is given by (\ref{leff_1}) or, better,  by its regularized form (\ref{lagB_reg}).
The second term
\begin{eqnarray}
\!\!\!\!\!\!\mathcal{L}_\mathrm{eff}(B,T)=
\sum_{\eta=\pm1}\sum_{j=1}^{\infty}\frac{(-)^{j+1}}{4\pi^2}
\int_0^\infty\frac{ds}{s^2}\int_0^\infty p_t\,dp_t
\exp\Bigl({\frac{j^2\beta^2}{4s}-s\Bigl(\sqrt{p_t^2+m^2}+\eta\frac{e\mu B}{4m}\Bigr)^2\Bigr)}\,,
\label{leff_2}
\end{eqnarray}
where the sum in $j$ has been arranged from one to infinity due to parity, represents
the contribution at non vanishing temperature. In view of the relation (\ref{Knu}) we can write
\begin{eqnarray}
&{}&\!\!\!\!\!\!\!\!\!\!\!\!\!\!\!\!\!\!\!\!\!\!\!\!\!\!\!\!\!\!\!\!\!\!\!\!\mathcal{L}_\mathrm{eff}(B,T)=\sum_{
\eta=\pm1}\sum_{j=1} ^{\infty} \frac{(-)^{j+1} }{\beta\pi^2 }
\int_0^\infty p_t\,dp_t\,K_1\Bigl[j\beta\Bigl(\sqrt{p_t^2+m^2}+\eta\frac{e\mu B}{4m}\Bigr)\Bigr]\cr
&{}&\qquad\qquad\qquad\qquad\qquad\qquad\qquad\qquad\cdot\,\Bigl(\sqrt{p_t^2+m^2}+\eta\frac{e\mu B}{4m}\Bigr)
\end{eqnarray}
By using the representation (\ref{K1}) of the Bessel function, we can give the final form for the temperature part of
free energy density of the spinor with a.m.m. in a magnetostatic field, namely
\begin{eqnarray}
\!\!\!\mathit{f}(B,T)=\frac{(-1)}{2\pi^2\beta}\sum_{\eta=\pm1}\int_0^{\infty}\!\! p_t\,dp_t\int_{-\infty}^{\infty}
\!\!dp_\ell\,
\ln\Bigr[1+\exp\Bigl(-\beta\sqrt{p_\ell^2+\Bigl(\sqrt{p_t^2+m^2}+\eta\frac{e\mu B}{4m}\Bigr)^2}\,\Bigr)\Bigr]
\label{fB}
\end{eqnarray}

%============================================================================================================
\medskip
\textbf{(}\textbf{\emph{ii}}\textbf{)} \textbf{\emph{The free energy for a spinor with a.m.m. in an electrostatic
field.}}
We now start from (\ref{Leff_es}) and make the usual substitutions described above, arriving at
\begin{eqnarray}
\mathcal{L}_\mathrm{eff}= \mathcal{L}_\mathrm{eff}(E,0)+\mathcal{L}_\mathrm{eff}(E,T)
\end{eqnarray}
where $\mathcal{L}_\mathrm{eff}(E,0)$ is given by the regularized expression (\ref{lagE_reg}) and
\begin{eqnarray}
&{}&\!\!\!\!\!\!\!\!\!\!\!\!\!\!\!\!\!\!\!\!\!\!\!\!\!\!\!\!\!\!\!\!\!\!\!\!\mathcal{L}
_\mathrm{eff}(E,T)=\sum_{
\eta=\pm1}\sum_{j=1} ^{\infty} \frac{(-)^{j+1} }{\beta\pi^2 }
\int_0^\infty p_t\,dp_t\,K_1\Bigl[j\beta\sqrt{m^2+\Bigl(p_t+\eta\frac{e\mu E}{4m}\Bigr)^2}\,\Bigr]\cr
&{}&\qquad\qquad\qquad\qquad\qquad\qquad\qquad\qquad\cdot\,\sqrt{m^2+\Bigl(p_t+\eta\frac{e\mu E}{4m}\Bigr)^2} 
\end{eqnarray}
Using once again (\ref{Knu})-(\ref{quattronove}), we eventually get 
\begin{eqnarray}
\!\!\!\mathit{f}(E,T)=-\frac1{2\pi^2\beta}\sum_{\eta=\pm1}\int_0^{\infty}\!\! p_t\,dp_t\int_{-\infty}^{\infty}
\!\!dp_\ell\,
\ln\Bigr[1+\exp\Bigl(-\beta\sqrt{p_\ell^2+m^2+\Bigl(p_t+\eta\frac{e\mu E}{4m}\Bigr)^2}\,\,\Bigr)\Bigr]
\label{fE} 
\end{eqnarray}
which represents the counterpart of (\ref{fB}) for the electrostatic case.

To conclude we would observe that in this paper we made a concrete use of a series of our previous results 
\cite{BGJP,BGPJP,BGPEP} in which we pointed out that the Foldy-Wouthuysen transformations can be used as
an efficient calculation tool for some physical systems. We previously proved \cite{BGPJP} that these are the cases in
which the FWT can be done exactly. In this paper  we have then applied the method to the case of relativistic
fermions with anomalous magnetic moment in uniform magnetostatic and electrostatic fields. We have thus been able to
produce a closed form for the effective action both at zero and finite temperature and therefore the free
Helmholtz energy of such systems.

\end{document}